\documentclass[iop,apj,times]{emulateapj}

\usepackage{graphicx}
\usepackage{apjfonts}

\slugcomment{A version adapted for publication in ApJ}

\shortauthors{Fukui et al.}

\begin{document}

\title{Multi-band, Multi-epoch Observations of the Transiting Warm Jupiter WASP-80b}


\author{Akihiko~Fukui\altaffilmark{1}, Yui~Kawashima\altaffilmark{2}, Masahiro~Ikoma\altaffilmark{2}, Norio~Narita\altaffilmark{3,4}, Masahiro~Onitsuka\altaffilmark{4}, Yoshifusa~Ita\altaffilmark{5}, Hiroki~Onozato\altaffilmark{5}, Shogo~Nishiyama\altaffilmark{3,6}, Haruka~Baba\altaffilmark{4}, Tsuguru~Ryu\altaffilmark{4}, Teruyuki~Hirano\altaffilmark{7}, Yasunori~Hori\altaffilmark{8,9}, Kenji~Kurosaki\altaffilmark{2}, Kiyoe~Kawauchi\altaffilmark{7}, Yasuhiro~H.~Takahashi\altaffilmark{3,10},  Takahiro~Nagayama\altaffilmark{11}, Motohide~Tamura\altaffilmark{12}, Nobuyuki~Kawai\altaffilmark{13}, Daisuke~Kuroda\altaffilmark{1}, Shogo~Nagayama\altaffilmark{3},  Kouji~Ohta\altaffilmark{14}, Yasuhiro~Shimizu\altaffilmark{1}, Kenshi~Yanagisawa\altaffilmark{1},  Michitoshi~Yoshida\altaffilmark{15}, and Hideyuki~Izumiura\altaffilmark{1}}

\altaffiltext{1}{Okayama Astrophysical Observatory, National Astronomical Observatory of Japan, Asakuchi, Okayama 719-0232, Japan; afukui@oao.nao.ac.jp}
\altaffiltext{2}{Department of Earth and Planetary Science, Graduate School of Science, The University of Tokyo, 7-3-1 Bunkyo-ku, Tokyo 113-0033, Japan}
\altaffiltext{3}{National Astronomical Observatory of Japan, 2-21-1 Osawa, Mitaka, Tokyo 181-8588, Japan}
\altaffiltext{4}{The Graduate University for Advanced Studies, 2-21-1 Osawa, Mitaka, Tokyo 181-8588, Japan}
\altaffiltext{5}{Astronomical Institute, Graduate School of Science, Tohoku University, 6-3 Aramaki Aoba, Aoba-ku, Sendai, Miyagi 980-8578, Japan}
\altaffiltext{6}{Faculty of Education, Miyagi University of Education, Sendai, 980-0845, Japan}
\altaffiltext{7}{Department of Earth and Planetary Sciences, Tokyo Institute of Technology, 2-12-1 Ookayama, Meguro-ku, Tokyo 152-8551, Japan}
\altaffiltext{8}{Department of Astronomy and Astrophysics, University of California, Santa Cruz, CA 95064, USA}
\altaffiltext{9}{Division of Theoretical Astronomy, National Astronomical Observatory of Japan, Osawa, Mitaka, Tokyo 1818588, Japan}
\altaffiltext{10}{Ministry of Education, Culture, Sports, Science and Technology, Japan, 3-2-2, Kasumigaseki, Chiyoda-ku, Tokyo 100-8959, Japan}
\altaffiltext{11}{Department of Physics, Nagoya University, Furo-cho, Chikusa-ku,Nagoya 464-8602, Japan}
\altaffiltext{12}{Department of Astronomy, Graduate School of Science, The University of Tokyo, and National Astronomical Observatory of Japan}
\altaffiltext{13}{Department of Physics, Tokyo Institute of Technology, 2-12-1, Oookayama, Meguro, Tokyo, 152-8551, Japan}
\altaffiltext{14}{Department of Astronomy, Kyoto University, Kitashirakawa-Oiwake, Sakyo, Kyoto, 606-8502, Japan}
\altaffiltext{15}{Hiroshima Astrophysical Science Center, Hiroshima University 1-3-1, Kagamiyama, Higashi-Hiroshima, Hiroshima, 739-8526, Japan}

\begin{abstract}
WASP-80b is a warm Jupiter transiting a bright late-K/early-M dwarf, providing a good opportunity to extend the atmospheric study of hot Jupiters toward the lower temperature regime. We report multi-band, multi-epoch transit observations of WASP-80b by using three ground-based telescopes covering from optical ($g'$, $R_\mathrm{c}$, and $I_\mathrm{c}$ bands) to near-infrared (NIR; $J$, $H$, and $K_\mathrm{s}$ bands) wavelengths. We observe 5 primary transits, each of which in 3 or 4 different bands simultaneously, obtaining 17 independent transit light curves.
Combining them with results from previous works, we find that the observed transmission spectrum is largely consistent with both a solar abundance and thick cloud atmospheric models at 1.7$\sigma$ discrepancy level. On the other hand, we find a marginal spectral rise in optical region compared to the NIR region at the 2.9$\sigma$ level, which possibly indicates the existence of haze in the atmosphere. We simulate theoretical transmission spectra for a solar abundance but hazy atmosphere, finding that a model with equilibrium temperature of 600~K can explain the observed data well, having a discrepancy level of 1.0$\sigma$. We also search for transit timing variations, but find no timing excess larger than 50 s from a linear ephemeris. In addition, we conduct 43 day long photometric monitoring of the host star in the optical bands, finding no significant variation in the stellar brightness. Combined with the fact that no spot-crossing event is observed in the five transits, our results confirm previous findings that the host star appears quiet for spot activities, despite the indications of strong chromospheric activities. 
\end{abstract}


\section{Introduction}
\label{sec:intro}

It is now well known that exoplanets have very different orbits reflecting diverse planetary origins and various migration mechanisms. To explore such planetary formation histories, unveiling their atmospheric compositions is important, because they can be affected by the environments in which they were born. For example, the relative abundances of carbon- and oxygen-bearing molecules in planetary atmospheres are closely related to the environment of surrounding disk gas \citep[e.g.,][]{2011ApJ...743L..16O,2012ApJ...758...36M}.

If an exoplanet has a transiting orbit, the planetary atmospheric composition can be examined by measuring transit and secondary-eclipse depths as a function of wavelength, which are referred to as transmission and emission spectroscopy, respectively. 
So far, several molecular features such as H$_2$O, CH$_4$, CO$_2$, and CO have been detected through these techniques in several hot Jupiters \citep[e.g.,][]{2007ApJ...661L.191B,2007Natur.448..169T,2010ApJ...712L.139T,2008Natur.452..329S, 2009ApJ...704.1616S,2008Natur.456..767G,2010Natur.463..637S,2010Natur.465.1049S,2013MNRAS.435.3481W,2013ApJ...774...95D},
although some of the detections are still in dispute \citep[e.g.][]{2009A&A...505..891S,2011MNRAS.411.2199G,2014ApJ...784..133S}.
On the other hand, flat or featureless transmission spectra have also been observed in a number of exoplanets, ranging from hot Jupiters, e.g., HD~189733b, WASP-12b, HAT-P-32b, and HAT-P-1b \citep{2008MNRAS.385..109P,2013MNRAS.436.2956S,2013MNRAS.436.2974G,2014MNRAS.437...46N} to super-Earths/Neptunes, e.g., GJ1214b and GJ436b \citep{2014Natur.505...69K,2014Natur.505...66K}, raising the possibility of the existence of floating small particles (aerosols) in their atmospheres as strong opacity sources. Such aerosols obscure or interfere with other atmospheric features \citep{2005MNRAS.364..649F}, preventing us from properly measuring the relative abundances of fundamental molecules such as H$_2$O, CH$_4$, CO$_2$, and CO. Therefore, understanding the nature of aerosols is crucial to explore planetary formation histories through their atmospheres.

Understanding the behaviors of aerosols is more important for cooler atmospheres. In an atmosphere with temperature below $\sim$1000~K, CH$_4$ becomes a major carbon carrier instead of CO. When such a CH$_4$-rich atmosphere is irradiated by UV fluxes, hydro-carbon haze particles (tholins) can be produced via photochemical processes. Indeed, such tholin hazes can be seen in CH$_4$-existing planets and satellites in our solar system (e.g., Uranus and Titan). However, few exoplanets with low-temperature atmospheres have been investigated to date. This is notable in gas giants\footnote{We note that atmospheres of low-temperature gas giants in wide orbits have recently been investigated by direct imaging \citep[e.g.,][]{2013ApJ...774...11K}.}, because transiting gas giants around late-type (low-temperature) stars are rare \citep[e.g.,][]{2012AJ....143..111J}.

In this context, a transiting Jovian planet WASP-80b was recently discovered around a nearby (60 $\pm$ 20 pc) late-K/early-M dwarf \citep{2013A&A...551A..80T}. This planet has several attractive points for studying its atmosphere. First, the planet has an equilibrium temperature of $\sim$800~K or less (assuming albedo is $>$0.1), which is one of the lowest temperatures among those of transiting Jovian planets around bright ($J<$10) host stars. Second, the host star shows strong chromospheric activities \citep{2013A&A...551A..80T,2014A&A...562A.126M}, possibly causing active photochemical reactions in the planetary atmosphere that might produce abundant tholin particles.
Last, the planet shows the second largest transit depth ($\sim$2.9 \%) following Kepler-45b ($\sim$3.2\%) among the known transiting planets, allowing us to measure the transit depth with relatively high precision. 
Therefore, WASP-80b offers a good opportunity to extend the atmospheric study toward the low-temperature regime, and test the photochemical reactions of hydro-carbons in an exoplanetary atmosphere.

Recently, \citet{2014A&A...562A.126M} conducted optical-to-near-infrared simultaneous observations for a transit of WASP-80b by using the GROND instrument on the MPG/ESO 2.2 m telescope ($g'$, $r'$, $i'$, $z'$, $J$, $H$, and $K$ bands) and the 1.54 m Danish telescope ($I$ band), reporting a flat transmission spectrum within the observational uncertainties. However, because the data are also consistent with a model spectrum assuming a solar-abundance atmosphere, further observations are needed to characterize its atmosphere. Furthermore, they omitted the $H$- and $K$-band observational data in their discussion of the spectrum due to relatively poor quality, leaving large room for improvement especially in the NIR regions.

In this paper, we report multi-band, multi-epoch transit observations of WASP-80b. We observed five primary transits, each of which in three or four different bands, using three ground-based telescopes at two observatories. Such multi-band and multi-epoch observations are quite useful not only to reduce statistical uncertainties but also to check systematics in measured parameters, especially for ground-based observations which often suffer from unknown systematics and/or lack of full-transit coverages.
We also report 43 day long photometric observations of the host star WASP-80 in three optical bands, to monitor its intrinsic variability. This is important because if the stellar brightness significantly varies from transit to transit due to such as star-spot appearing/vanishing, the observed transmission spectrum, or practically star-to-planet radius ratio $R_\mathrm{p}/R_\mathrm{s}$, can be biased \citep[e.g.][]{2008MNRAS.385..109P}.

The remainder of this paper is organized as follows. We describe our observations in Section 2. The methods of reduction and analysis of the observed data are shown in Section 3. The atmospheric properties, transit timings, and stellar variability of the WASP-80 system are discussed in Section 4. Finally, we summarize this paper in Section 5.

\section{Observations}

\subsection{Transit Observations with IRSF/SIRIUS}
\begin{figure*}
\begin{center}
\includegraphics[width=14cm]{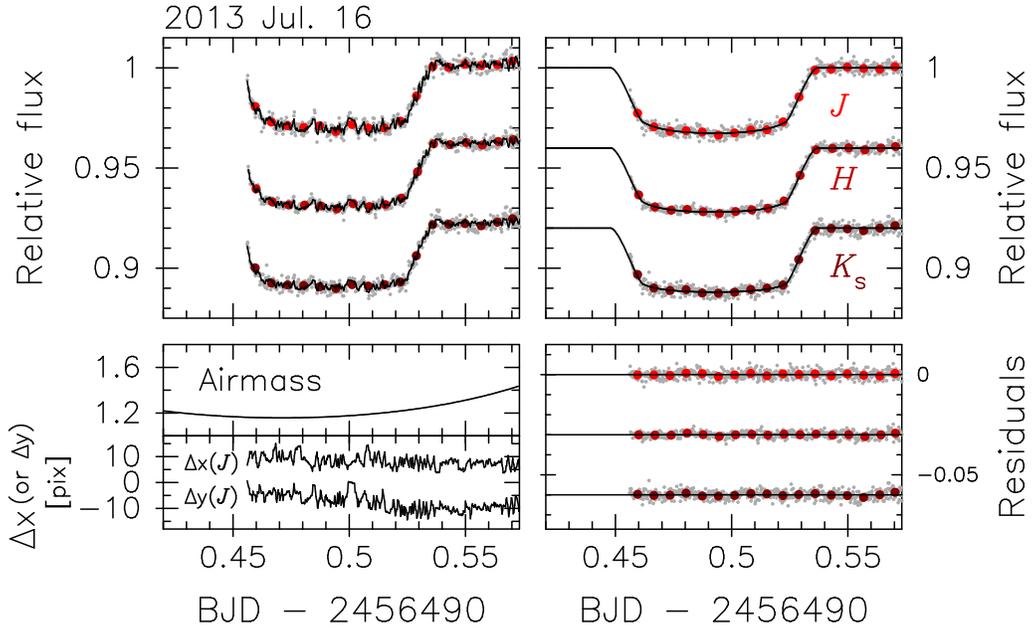}
\caption{(Left upper) Baseline-uncorrected transit light curves of WASP-80b observed with IRSF/SIRIUS on 2013 July 16. The $J$-, $H$-, and $K_\mathrm{s}$-band light curves are shown from top to bottom, where the lower two are arbitrary shifted along the vertical direction for display. The small gray points and large colored points are unbinned and 10 minute binned data, respectively. The solid lines are the best-fit transit+baseline models derived from the MCMC analysis. (Left lower) The air mass change (top), and the stellar positional changes in $x$ and $y$ directions on the $J$-band detector, where arbitrary constant values are added for clarity (bottom). (Right upper) Baseline-corrected light curves. (Right lower) Residual light curves.\label{fig:lc-130716}}
\end{center}
\end{figure*}

\begin{figure*}
\begin{center}
\includegraphics[width=14cm]{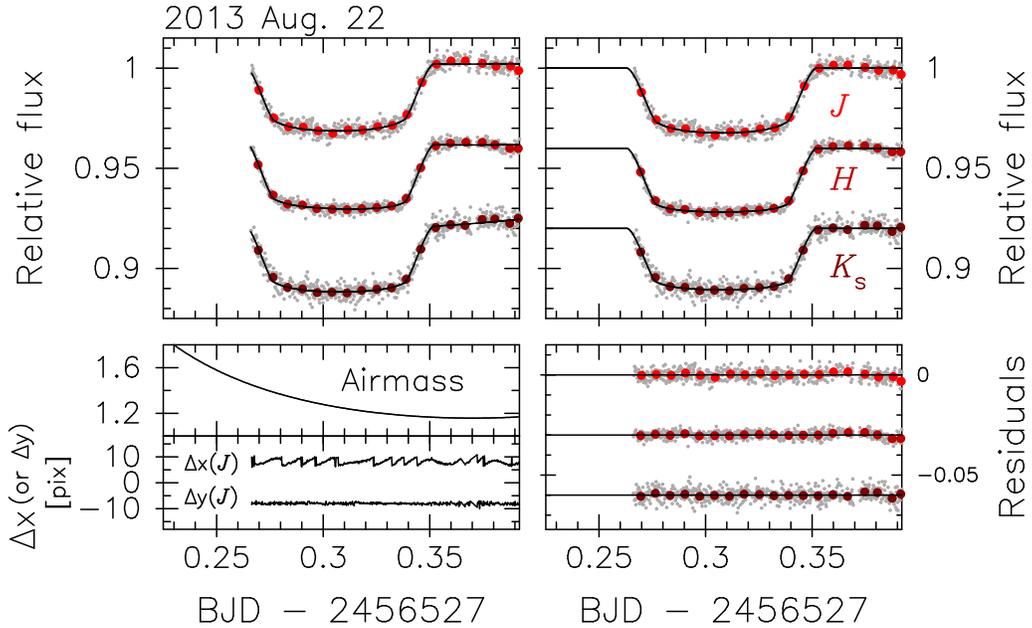}
\caption{Same as Figure \ref{fig:lc-130716}, but for the data obtained on 2013 August 22. \label{fig:lc-130822}}
\end{center}
\end{figure*}

\begin{figure*}
\begin{center}
\includegraphics[width=14cm]{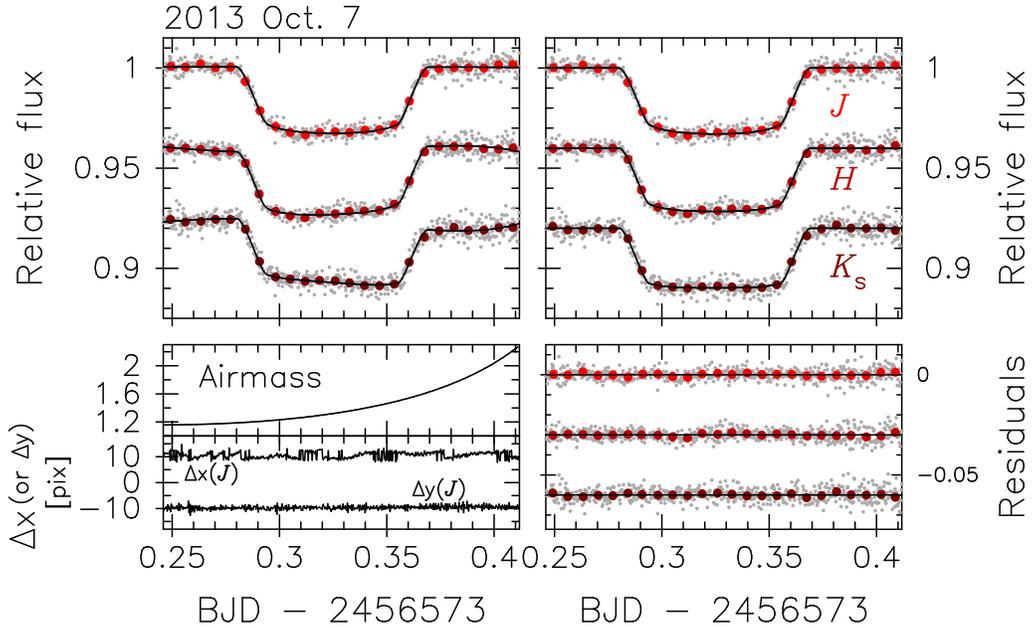}
\caption{Same as Figure \ref{fig:lc-130716}, but for the data obtained on 2013 October 7. \label{fig:lc-131007}}
\end{center}
\end{figure*}

We observed three primary transits of WASP-80b with the Simultaneous Infrared Imager for Unbiased Survey \citep[SIRIUS;][]{2003SPIE.4841..459N} camera mounted on the Infrared Survey Facility (IRSF) 1.4~m telescope at the South African Astronomical Observatory on 2013 July 16, August 22, and October 7 (UT). SIRIUS has three detectors, each consisting of 1k $\times$ 1k pixels with a pixel scale of 0\farcs45 pixel$^{-1}$, enabling us to obtain $J$-, $H$-, and $K_s$-band images simultaneously with a field of view (FOV) of 7\farcm7 $\times$ 7\farcm7. 
During each observation, we defocused stellar images so that the FWHM of stellar point-spread function (PSF) was 14--17 pixels on July 16 and October 7, and 9.5--12 pixels on  August 22, in order to improve photometric precision. In addition, we activated a software that corrects tracking errors by calculating the stellar positional shift on the newly obtained $J$-band image and feeding it back to the telescope. 
The exposure time was set to 10 s on July 16 and August 22, and 15 s on October 7.
The weather was photometric without any thin cloud passing for the three nights. We observed full transit including pre- and post-transit parts on July 16 and October 7; however, we discarded the data before 22:47 on July 16 (UT) because of an accidental drift of the stellar position on the detector that causes uncorrectable systematic errors on photometry. The pre-transit part on August 22 was not observed due to interference with another observational program.
In the lower left panel in Figures \ref{fig:lc-130716}--\ref{fig:lc-131007}, we show the air-mass change (top) and stellar positional change along $x$ and $y$ directions on the $J$-band detector (bottom) during the respective observations.  An observing log is shown in Table \ref{tbl:obslog}.

\begin{deluxetable}{lcccc}
\tablecaption{Observing Log\label{tbl:obslog}}
\tablehead{
Date &  \colhead{Telescope/} & \colhead{Filter} & \colhead{Exp. Time} & \colhead{$N_\mathrm{obs}$ $^{a}$}\\
(UT)& \colhead{Instrument} & & (s) & 
}
\startdata
2013 Jul 16  &  IRSF/SIRIUS & $J$ & 10 & 541\\[1pt]
 & IRSF/SIRIUS & $H$ & 10 & 537 \\[1pt]
 & IRSF/SIRIUS  & $K_\mathrm{s}$ & 10 & 539\\[2pt]
\hline\\[-5pt]
2013 Aug 13 &  MITSuME 50cm & $g'$ & 30 & 562 \\[1pt]
 & MITSuME 50cm &  $R_\mathrm{c}$ & 30 & 566 \\[1pt]
 & MITSuME 50cm & $I_\mathrm{c}$ & 30 & 563 \\[1pt]
 & OAO188cm/ISLE & $J$ & 45 & 324 \\[2pt]
\hline\\[-5pt]
2013 Aug 22 & IRSF/SIRIUS & $J$ &  10 & 566 \\[1pt]
 & IRSF/SIRIUS & $H$ & 10 & 576 \\[1pt]
 & IRSF/SIRIUS & $K_\mathrm{s}$ & 10 & 582 \\[2pt]
\hline\\[-5pt]
2013 Sep 22 & MITSuME 50cm & $g'$ & 30 & 268 \\[1pt]
 & MITSuME 50cm & $R_\mathrm{c}$ & 30 & 290 \\[1pt]
 & MITSuME 50cm & $I_\mathrm{c}$ & 30 & 295 \\[1pt]
 & OAO188cm/ISLE & $J$ & 45 & 241 \\[2pt]
\hline\\[-5pt]
2013 Oct 7 & IRSF/SIRIUS & $J$ &  15 & 597 \\[1pt]
 & IRSF/SIRIUS & $H$ & 15 & 597 \\[1pt]
 & IRSF/SIRIUS & $K_\mathrm{s}$ & 15 & 603
\enddata
\tablenotetext{a}{\ The number of observed data points excluding outliers.}
\end{deluxetable}

\subsection{Transit Observations with OAO188cm/ISLE and MITSuME}
We observed two primary transits of WASP-80b by simultaneously using two instruments at Okayama Astrophysical Observatory (OAO) on 2013 August 13 and September 22  (UT); one is the NIR imaging and spectroscopic instrument ISLE \citep{2006SPIE.6269E.118Y,2008SPIE.7014E.106Y} mounted on the 188~cm telescope and the other is a multi-color imager mounted on the 50~cm telescope which is one of Multicolor Imaging Telescopes for Survey and Monstrous Explosions \citep[MITSuME;][]{2005NCimC..28..755K,2010AIPC.1279..466Y}. 

ISLE has a 1k $\times$ 1k HAWAII-1 array having a pixel scale of 0\farcs25 pixel$^{-1}$ and a FOV of 4\farcm5 on a side.
We used $J$-band filter and set the exposure time to 45 s on both nights. We defocused stellar images so that the FWHM of stellar PSF was 23--27 pixels. In addition, we activated a hybrid auto-guiding system \citep{2013ApJ...770...95F}, which consists of an off-axis auto-guiding camera to correct telescope's tracking errors in real time, and a software that corrects a gradual drift of the origin of the auto-guiding camera with respect to the ISLE detector by calculating the shift of stellar positions on the ISLE images.

The multi-color imager mounted on the MITSuME telescope consists of three 1k $\times$ 1k CCDs, enabling us to obtain $g'$-, $R_\mathrm{c}$-, and $I_\mathrm{c}$-band images simultaneously. Each CCD has a pixel scale of 1\farcs5 pixel$^{-1}$, providing a FOV of 26$'$ on a side. We slightly defocused the stellar images so that the FWHM of PSF was 1.2--2.5 pixels. We also activated a software that corrects the stellar positional shift on the  $I_\mathrm{c}$-band detector soon after each exposure. The exposure time was set to 30 s on both nights.

\begin{figure*}
\begin{center}
\includegraphics[width=14cm]{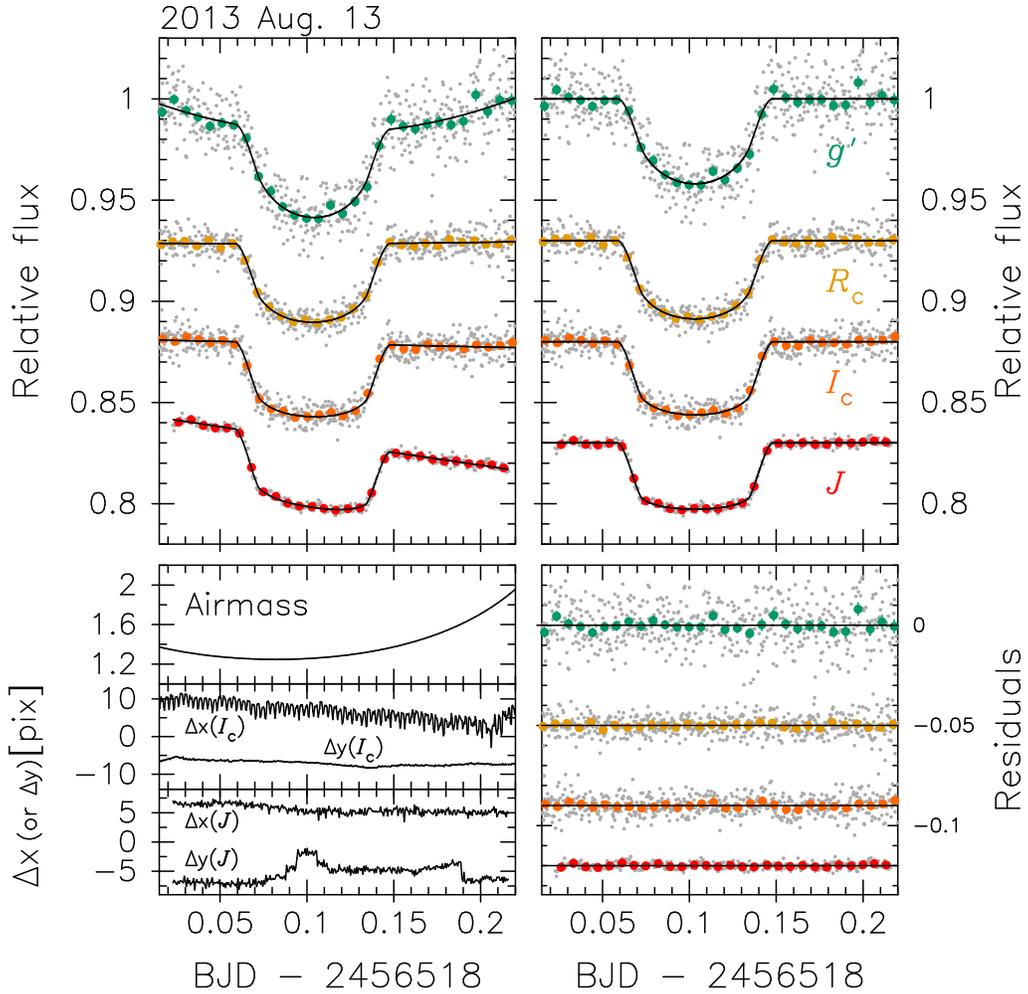}
\caption{(Left upper) Baseline-uncorrected transit light curves of WASP-80b observed at OAO on 2013 August 13.  The $g'$-, $R_\mathrm{c}$-, $I_\mathrm{c}$-, and $J$-band light curves are shown from top to bottom, where the lower three are arbitrary shifted along the vertical direction for display. The other meanings are the same as in Figure \ref{fig:lc-130716}. (Left lower) The air mass change (top), and the stellar positional changes in $x$ and $y$ directions on the MITSuME/$I_\mathrm{c}$-band detector (middle) and those on the ISLE/$J$-band detector (bottom), where arbitral constant values are added for clarity. (Right upper) Baseline-corrected light curves. (Right lower) Residual light curves.\label{fig:lc-130813}}
\end{center}
\end{figure*}

\begin{figure*}
\begin{center}
\includegraphics[width=14cm]{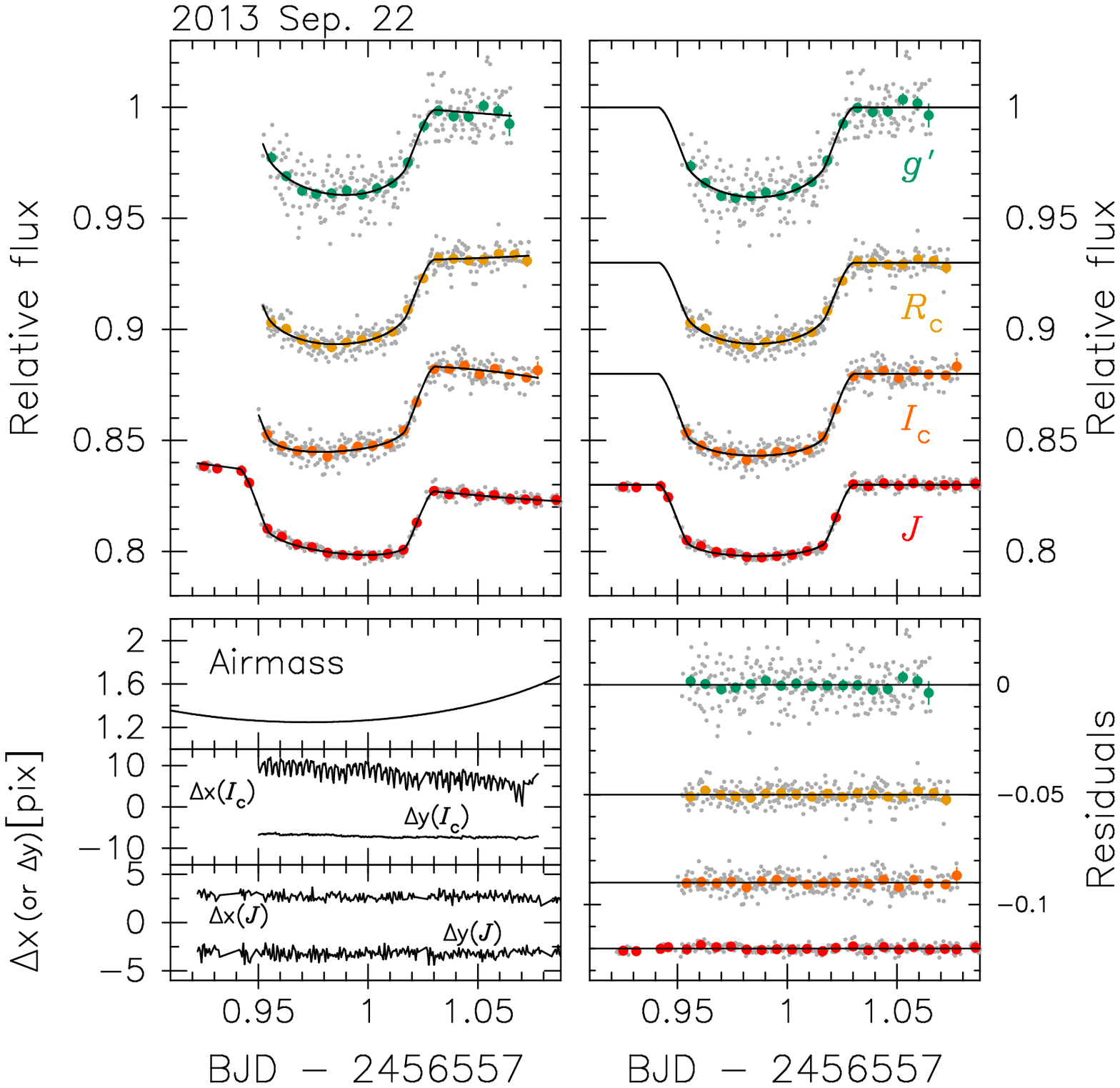}
\caption{Same as Figure \ref{fig:lc-130813}, but for the data obtained on 2013 September 22. \label{fig:lc-130922}}
\end{center}
\end{figure*}

The weather was mostly photometric on both nights, while some thin clouds passed at the beginning of the observation on September 22, which results in large flux drops especially in the optical bands; we discarded these data from the analysis in the remainder of this paper. 
In the lower left panel in Figure \ref{fig:lc-130813} and \ref{fig:lc-130922}, we show air-mass change (top) and stellar positional change along $x$ and $y$ directions on the ISLE (middle) and MITSuME/$I_\mathrm{c}$-band (bottom) detectors during the respective observations.
An observing log is compiled in Table \ref{tbl:obslog}.

\subsection{Photometric Monitoring of Stellar Variability with MITSuME}
In order to check the intrinsic variability of the host star WASP-80 around the period of our transit observations, we conducted out-of-transit observations on 14 nights spanning 43 days from 2013 August 10 to 2013 September 22, by using the 50 cm MITSuME telescope in $g'$, $I_\mathrm{c}$, and $R_\mathrm{c}$ bands. All the settings were the same as the transit observations described in the previous section. The observations were conducted for one to two hours on each night, and in total about 1700 images were gathered for each band.

\section{Analysis}
\subsection{Data Reduction}
\label{sec:reduction}
All the observed images are dark-subtracted and flat-fielded in a standard manner. The flat-field images are created from dozens of twilight flat images that were obtained before and after each observation for the SIRIUS and MITSuME data, and from 100 dome-flat images that were taken on each observing night for the ISLE data. After that, aperture photometry is performed for  the target and several (for SIRIUS and ISLE) or dozens (for MITSuME) of bright stars spread on the reduced images, by using a customized tool with constant-aperture-radius mode \citep{2011PASJ...63..287F}. The target flux is divided by the sum of the fluxes of a selected number of bright stars (comparison stars) to produce a relative light curve.

The time for each data point is assigned as the mid-time of exposure in the Barycentric Julian Day (BJD) time system based on Barycentric Dynamical Time (TDB), which is converted from Julian Day (JD) based on Coordinated Universal Time (UTC), recorded on the FITS header, via the code of \citet{2010PASP..122..935E}.

\subsection{Preparation of Transit Light Curves}
\label{sec:prepare_lc}
In order to optimize the set of comparison stars and aperture radius for each instrument, filter, and transit (each data set), we produce a number of trial light curves for each data set by changing the combination of comparison stars, as well as changing the aperture radius with a step size of 0.5 pixel (for MITSuME) or 1 pixel (for others).
Then, we fit the individual trial light curves with a transit-plus-baseline model to select the best light curve so that the  root mean square (rms) of the residual light curve is minimum. For the light curve model, we use the following functions:
\begin{eqnarray}
F&=& k_0 \times 10^{-0.4 \Delta m_\mathrm{corr}} \times F_\mathrm{tr},\\
\Delta m_\mathrm{corr} &=& \sum_{i=1} k_i X_i,
\end{eqnarray}
where $F$ is the relative flux, $F_\mathrm{tr}$ is the transit light-curve model,  $\{{\bf X}\}$ are variables for the baseline function, and $\{{\bf k}\}$ are coefficients.
For the variables $\{{\bf X}\}$, we tentatively use \{$t$, $z$\}, where  
$t$ is time and $z$ is air mass.
For the transit model $F_\mathrm{tr}$, we use the analytic formula given by \citet{2009ApJ...690....1O}, which is equivalent to that of \citet{2002ApJ...580L.171M} when using the quadratic limb-darkening raw. 
The transit parameters we use are the mid-transit time $T_\mathrm{c}$,  the planet-star radius ratio $R_p/R_s$, the semi-major axis normalized by the stellar radius $a/R_s$, 
 the orbital inclination $i_\mathrm{orb}$, 
and the quadratic limb-darkening coefficients $u_1$ and $u_2$.  
Among these parameters, $T_\mathrm{c}$, $R_p/R_s$ and $u_1$ are let free, while $a/R_s$ and $i_\mathrm{orb}$ are fixed at the values derived from \citet{2014A&A...562A.126M}, namely, 12.6119 and 88.91 deg, respectively. We also fix $u_2$ at the theoretical values for a star with log$g$=4.5 and $T_\mathrm{eff}$=4100 K for respective filters, adopted from \citet{2012A&A...546A..14C}, namely, 0.109, 0.191, 0.225, 0.223, 0.267, 0.241 for $g'$, $R_\mathrm{c}$, $I_\mathrm{c}$, $J$, $H$, and $K_s$, respectively.
A circular orbit, with an orbital period of $P = 3.06786144$ days adopted from \citet{2014A&A...562A.126M}, is assumed.
The individual light curves are fitted by the AMOEBA algorithm \citep{1992nrca.book.....P} to find the one that gives the minimum rms value by iteratively eliminating $>$4$\sigma$ outliers. 
In Table \ref{tbl:prepare_lc}, we summarize the number of selected comparison stars and the selected aperture radius for all data sets. The selected light curves are shown in the upper left panels in Figures \ref{fig:lc-130716}--\ref{fig:lc-130922}, where 10 minute binned data are also shown as a visual guide. 
We note that there exists a fainter neighboring star ($\Delta B$=2.8 and $\Delta K_\mathrm{s}$=4.0) that is separated from WASP-80 by 8\farcs9; we confirm that with the selected aperture radii the flux contamination from the fainter star is negligible for all the data sets.

\subsection{Selection of Baseline Models}
After preparing the light curves, we select the best-describing baseline model, i.e., which variables should be included in  $\{{\bf X}\}$ in Equation (2), for each light curve.
To do so, we first fit each light curve with Equations (1) and (2), letting $T_\mathrm{c}$, $R_p/R_s$, $u_1$, and $a/R_\mathrm{s}$ be free while fixing others, by changing the set of variables $\{{\bf X}\}$. The set of variables are chosen from $t$, $z$, $t^2$, $\Delta x$, and $\Delta y$, where $\Delta x$ and $\Delta y$ are the relative stellar displacement along the $x$ and $y$ directions, respectively, on the detectors.
Next, we evaluate the Bayesian information criteria \citep[BIC;][]{1978Schwarz} for each baseline model; the BIC value is given by BIC=$\chi^2 + k \ln N$, where $k$ is the number of free parameters and $N$ is the number of data points. Finally, we select the best baseline model such that the BIC value is minimum.

\begin{figure*}
\begin{center}
\includegraphics[width=16cm]{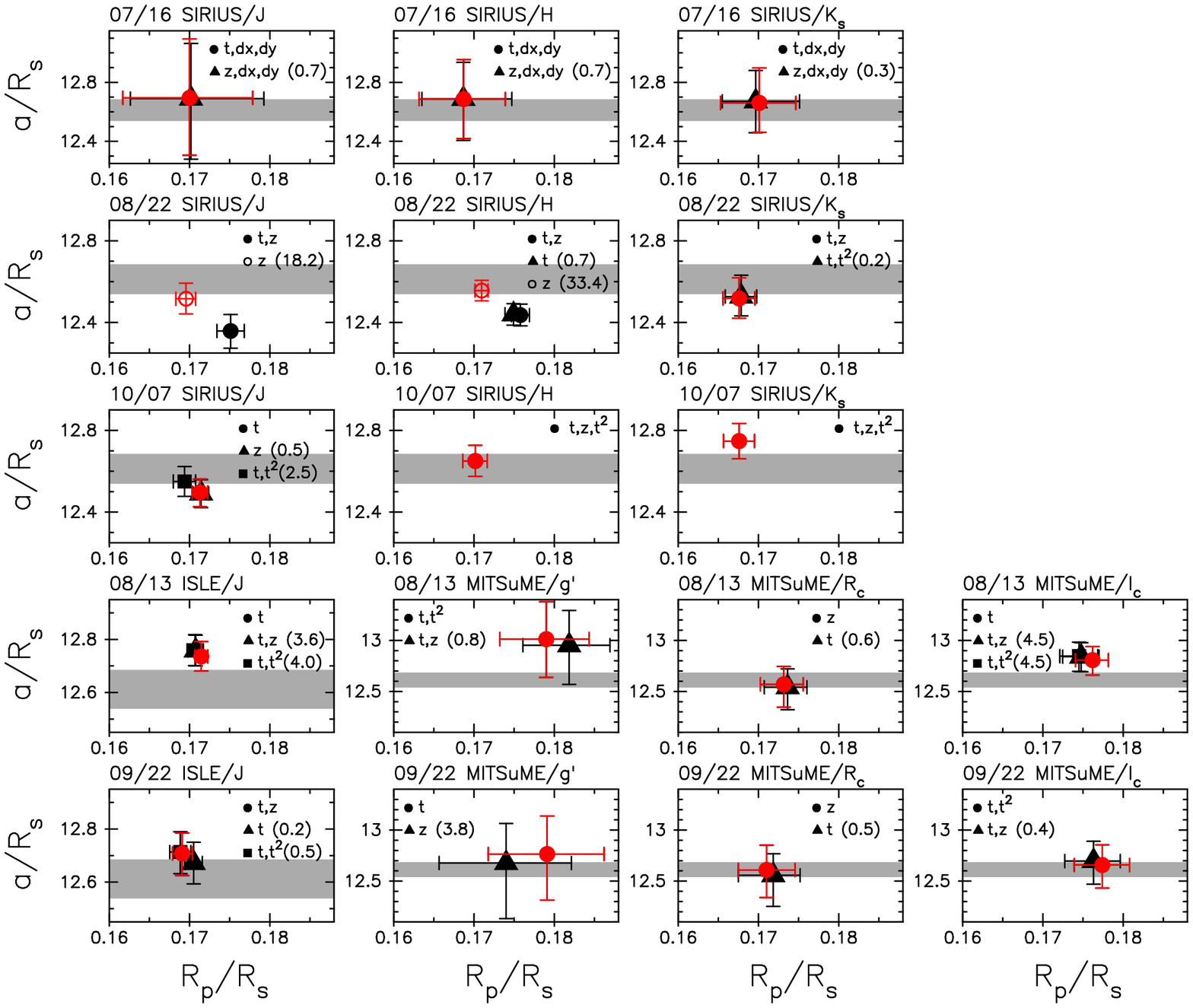}
\caption{Comparison of $R_\mathrm{p}/R_\mathrm{s}$ and $a/R_\mathrm{s}$ derived by the AMOEBA fitting for each light curve using different baseline models. The results for each light curve are shown in each panel, on top of which the observing month/day and instrument/filter are indicated. The filled circle, triangle, and square denote the values derived by using the baseline models that give the minimum, second minimum, and third minimum BIC values, respectively. The number in parenthesis in the legend indicates $\Delta$BIC with respect to the minimum BIC value. Note that only the plots giving $\Delta$BIC$<$5 are shown in this figure, except for the panels for  SIRIUS/$J$ and $H$ on August 22, where the values derived from \{{\bf X}\}  = \{$t$\} are additionally shown as open circles. The gray belt indicates the 1$\sigma$ credible region of $a/R_\mathrm{s}$ derived by \citet{2014A&A...562A.126M}. Our choices for the best-describing baseline models are indicated by red color.
\label{fig:RpRs_vs_aRs}}
\end{center}
\end{figure*}

This procedure works well for all the light curves except for the $J$- and $H$-band light curves obtained on 2013 August 22.
For these two light curves, we find that the fittings with the minimum-BIC models, namely $\{{\bf X}\}$=\{$t, t^2$\} for both, derive inconsistent $a/R_\mathrm{s}$ values with that derived from \citet{2014A&A...562A.126M}, although for all the other light curves the comparable fittings derive consistent $a/R_\mathrm{s}$ values, largely within their uncertainties (see Figure \ref{fig:RpRs_vs_aRs}). The two exceptional light curves lack a pre-transit part; in such a case, incorrect baseline models can incidentally fit the data well. If this is the case, the derived $R_\mathrm{p}/R_\mathrm{s}$ value, which is what we want to measure, could be shifted from the true value. In fact, the respective $R_\mathrm{p}/R_\mathrm{s}$ values for the $J$- and $H$-band light curves on August 22 are significantly larger than those for the light curves from the same band on October 7 (see the corresponding panels in Figure  \ref{fig:RpRs_vs_aRs}), which depict both before and after the transit.
For the two exceptions, we alternatively find that a simpler baseline model of $\{{\bf X}\}$ = \{$z$\} can fit them giving consistent $a/R_\mathrm{s}$ values with those from Mancini et al. as well as consistent $R_\mathrm{p}/R_\mathrm{s}$ values with the light curves from the same band on October 7 (denoted as open circles in Figure \ref{fig:RpRs_vs_aRs}), while the BIC differences between $\{{\bf X}\}$ = \{$z$\} and \{$t, t^2$\} for the $J$- and $H$-band light curves are 18.8 and 34.1, respectively. For the above reasons, we choose $\{{\bf X}\}$ = \{$z$\} as the best baseline model for these two light curves. We note that it is not likely that $a/R_\mathrm{s}$ and $R_\mathrm{p}/R_\mathrm{s}$ have changed over time, because fitting to the $K_\mathrm{s}$-band light curve on the same night with the minimum-BIC model gives a $a/R_\mathrm{s}$ value that is consistent with that of Mancini et al. 
In Table \ref{tbl:prepare_lc}, we summarize the selected sets of variables \{{\bf X}\} for the respective light curves.

We also note that even for the other light curves, there can be several baseline models that give BIC values similar to the minimum one, potentially causing systematic errors on $R_\mathrm{p}/R_\mathrm{s}$ depending on which baseline model we select. To see the impact of this possibility, we also plot in Figure \ref{fig:RpRs_vs_aRs} the $R_\mathrm{p}/R_\mathrm{s}$ and  $a/R_\mathrm{s}$ values for the baseline models that give the BIC difference with respect to the minimum-BIC model ($\Delta$BIC) of less than five. As a result, the $R_\mathrm{p}/R_\mathrm{s}$ and  $a/R_\mathrm{s}$ values for similar-BIC models (filled circle, triangle, and square are for the minimum, second-, and third-minimum BIC models, respectively) are close to each other compared to their uncertainties, implying that the systematics on $R_\mathrm{p}/R_\mathrm{s}$ due to different baseline models are small.

After the baseline-selection process, we rescale the flux uncertainties in each light curve such that the reduced $\chi^2$ of the transit-plus-baseline-model fit becomes unity. In addition, we further rescale these uncertainties by the so-called $\beta$ factor \citep{2008ApJ...683.1076W}  taking red noises into account. The $\beta$ factor is defined as $\sigma_{N, \mathrm{obs}}$/$\sigma_{N, \mathrm{exp}}$, where $\sigma_\mathrm{N,obs}$ is the standard deviation of the residual light curve binned by $N$ data points into $M$ bins, and $\sigma_{N, \mathrm{exp}} \equiv \sigma_1 \sqrt{M/N(M-1)}$ is the expected standard deviation for the binned residual light curve assuming that the unbinned residuals with the standard deviation of $\sigma_1$ are dispersed in a Gaussian distribution. We take the median value of $\beta$ calculated for $N=$ 4 to 15 for each light curve. The calculated $\beta$ values are summarized in Table \ref{tbl:prepare_lc}.

\begin{deluxetable*}{lccccccc}
\tablewidth{9cm}
\tabletypesize{\footnotesize}
\tablecaption{Summary of Light Curve Preparation \label{tbl:prepare_lc}}
\tablehead{
Instrument & Obs. Date & Filter & $N_\mathrm{comp}$ $^{a}$ & $R_\mathrm{ap}$ $^{b}$ & $\{{\bf X}\}$ & RMS & $\beta$\\
& (UT) & & & (pixel) & & (\%) &
}
\startdata
SIRIUS & 2013 Jul 16 & $J$ & 3 & 12.0 & $t$, $\Delta x$, $\Delta y$ & 0.194 & 1.24\\[1pt] 
 & & $H$ & 3 & 14.0 & $t$, $\Delta x$, $\Delta y$ & 0.145 & 1.58\\[1pt] 
 & & $K_\mathrm{s}$ &  4 & 12.0 & $t$, $\Delta x$, $\Delta y$ & 0.200 & 1.33\\[1pt]
 & 2013 Aug 22 & $J$ & 3 & 9.0 & $z$ & 0.251 & 1.36\\[1pt] 
 & & $H$ & 3 & 9.0 & $z$ & 0.177 & 1.68\\[1pt] 
 & & $K_\mathrm{s}$ & 2 & 9.0 & $t$, $z$ & 0.298 & 1.13\\[1pt] 
 & 2013 Oct 7 & $J$ & 3 & 16.0 & $t$ & 0.270 & 1.06\\[1pt] 
 & & $H$ & 2 & 16.0 & $t$, $z$, $t^2$ & 0.265 & 1.17\\[1pt] 
 & & $K_\mathrm{s}$  & 3 & 14.0 & $t$, $z$, $t^2$ & 0.329 & 1.01\\[2pt] 
\hline\\[-5pt]
ISLE & 2013 Aug 13 & $J$ & 1 & 20.0 & $t$ & 0.164 & 1.35\\[1pt]
& 2013 Sep 22 & $J$ & 1 & 21.0 & $t$, $z$ & 0.177 & 1.28\\[2pt]
\hline\\[-5pt]
MITSuME & 2013 Aug 13 & $g'$ & 8 & 2.5 & $t$, $t^2$ & 1.05 & 1.02\\[1pt]
& & $R_\mathrm{c}$  & 6 & 4.0 & $z$ & 0.477 & 1.00\\[1pt]
& & $I_\mathrm{c}$ & 9 & 3.5 & $t$ & 0.491 & 1.01\\[1pt]
& 2013 Sep 22 & $g'$ & 8 & 3.0 & $t$ & 0.855 & 1.00\\[1pt]
& & $R_\mathrm{c}$  & 6 & 4.0 & $z$ & 0.417 &1.12\\[1pt]
& & $I_\mathrm{c}$ & 9 & 3.5 & $t$, $t^2$ & 0.404 & 1.08
\enddata
\tablenotetext{a}{\ The number of comparison stars.}
\tablenotetext{b}{\ Aperture radius.}
\end{deluxetable*}

\subsection{MCMC Analysis}
\label{sec:mcmc}

To properly derive the $R_\mathrm{p}/R_\mathrm{s}$ values and their uncertainties for the respective data sets,  we perform the Markov Chain Monte Carlo (MCMC) analysis for each transit by using a customized code \citep{2007PASJ...59..763N,2013PASJ...65...27N}.
In this analysis, all the light curves involved in one transit are analyzed simultaneously, treating $i_\mathrm{orb}$, $a/R_\mathrm{s}$, and $T_\mathrm{c}$ as common parameters, whereas $R_\mathrm{p}/R_\mathrm{s}$, $u_1$, $u_2$, and $\{{\bf k}\}$ are treated as independent parameters for the respective light curves. Here, $\{{\bf k}\}$ are the coefficients corresponding to the variables \{{\bf X}\} selected in the previous section. In the same way as in Section \ref{sec:prepare_lc},  we fix $i_\mathrm{orb}$ and $a/R_\mathrm{s}$ at the values from \citet{2014A&A...562A.126M}, and fix $u_2$ at the theoretical values for the respective filters; we let  the other adjustable parameters be free. 
The reason for fixing the $i_\mathrm{orb}$ and $a/R_\mathrm{s}$ values to those of Mancini et al. is that these parameters are correlated with $R_\mathrm{p}$/$R_\mathrm{s}$, and varying them would cause systematic offset on measured $R_\mathrm{p}$/$R_\mathrm{s}$, which is what we aim to compare between different bands among the data including the ones from Mancini et al. (see Section \ref{sec:spectrum}).

We start the MCMC procedure with the best-fit parameters determined by the AMOEBA algorithm, using their 1$\sigma$ uncertainties as the widths of Gaussian jump functions for updating MCMC steps. We perform 10 sequential MCMC runs with 10$^6$ chained steps in each run, updating the best-fit parameters and their 1$\sigma$ uncertainties. To wait for convergence, we discard the first five MCMC runs. The final median values and 1$\sigma$ uncertainties of the respective parameters are calculated from the merged posterior-probability distributions from the last five MCMC runs. 

The resultant parameters are summarized in Table \ref{tbl:mcmc}. The final light curve models are displayed as solid lines in the upper left panels in Figures \ref{fig:lc-130716}--\ref{fig:lc-130922}, as well as the baseline-corrected light curves and residual light curves are shown in the upper right and lower right panels, respectively, in the same figures.
We note that no apparent spot-crossing event is seen in any of the five transits.

\begin{deluxetable*}{lcccccc}
\tablewidth{18.4cm}
\tabletypesize{\scriptsize}
\tablecaption{MCMC Results$^\mathrm{a}$ \label{tbl:mcmc}}
\tablehead{
Parameter & & & Value & &\\
& \colhead{2013 Jul 16} & \colhead{2013 Aug 13} & \colhead{2013 Aug 22} & \colhead{2013 Sep 22} & \colhead{2013 Oct 7} & Mancini et al. (2014)
}
\startdata
$T_\mathrm{c}$ & 6490.492507  & 6518.10335  & 6527.30758  & 6557.98585  & 6573.32468 &... \\[1pt]
\ [BJD$_\mathrm{TDB}$-2450000]& \ \ \ $\pm$ 0.000091 & \ \ \ $\pm$ 0.00012 & \ \ \ $\pm$ 0.00016 & \ \ \ $\pm$ 0.00018 &  \ \ \ $\pm$ 0.000088 &\\[1pt]
%
%
$u_1$ ($g'$) & ... & 0.842 $\pm$ 0.087 & ... & 0.642 $^{0.095}_{-0.097}$ & ... & ...\\[1pt]
$u_1$ ($R_\mathrm{c}$) & ... & 0.624 $^{0.042}_{-0.043}$ & ... & 0.551 $^{0.056}_{-0.059}$ & ... &...\\[1pt]
$u_1$ ($I_\mathrm{c}$) & ... & 0.407 $^{0.048}_{-0.050}$ & ... & 0.353 $^{0.052}_{-0.054}$ & ... &... \\[1pt]
$u_1$ ($J$) & 0.270 $\pm$ 0.026 & 0.254 $\pm$ 0.032 & 0.265 $^{0.035}_{-0.037}$ & 0.258 $^{0.038}_{-0.040}$ & 0.204 $\pm$ 0.027&... \\[1pt]
$u_1$ ($H$) & 0.213 $\pm$ 0.023 & ... & 0.126 $^{0.033}_{-0.031}$ & ... & 0.133 $^{0.033}_{-0.038}$ &... \\[1pt]
$u_1$ ($K_\mathrm{s}$) & 0.153 $\pm$ 0.029 & ... & 0.159 $\pm$ 0.038 & ... & 0.048 $^{0.048}_{-0.053}$ &... \\[1pt]
$R_\mathrm{p}/R_\mathrm{s}$ ($g'$) & ... & 0.1743 $^{0.0045}_{-0.0047}$ & ... & 0.1787 $\pm$ 0.0064 & ... & 0.17033 $\pm$ 0.00217\\[1pt]
$R_\mathrm{p}/R_\mathrm{s}$ ($R_\mathrm{c}$) & ... & 0.1736 $\pm$ 0.0017 & ... & 0.1711 $^{0.0040}_{-0.0037}$ & ...&0.17041 $\pm$ 0.00175 \\[1pt]
$R_\mathrm{p}/R_\mathrm{s}$ ($I_\mathrm{c}$) & ... & 0.1741 $\pm$ 0.0015 & ... & 0.1778 $\pm$ 0.0039 & ... &0.17183 $\pm$ 0.00161 \\[1pt]
$R_\mathrm{p}/R_\mathrm{s}$ ($J$) & 0.1697 $\pm$ 0.0015 & 0.1704 $\pm$ 0.00092 & 0.1690 $\pm$ 0.0014 & 0.1686 $\pm$ 0.0016 & 0.17234 $^{0.00089}_{-0.00081}$ & 0.1695 $\pm$ 0.0028 \\[1pt]
$R_\mathrm{p}/R_\mathrm{s}$ ($H$) & 0.1688 $\pm$ 0.0014 & ... & 0.1709 $\pm$ 0.0012 & ... & 0.1702 $\pm$ 0.0017 &... \\[1pt]
$R_\mathrm{p}/R_\mathrm{s}$ ($K_\mathrm{s}$) & 0.1708 $\pm$ 0.0016 & ... & 0.1672 $\pm$ 0.0022 & ... & 0.1679 $\pm$ 0.0023 & ...
\enddata
\tablenotetext{a}{\  The MCMC analysis in this work is performed by fixing $i_\mathrm{orb} =$ 88.91 deg and $a/R_\mathrm{s}=$ 12.612, which are obtained from  \citet{2014A&A...562A.126M}.}
\end{deluxetable*}

\subsection{Photometric Monitoring Data}
\label{sec:longlc}

\begin{figure}
\begin{center}
\includegraphics[width=8.5cm]{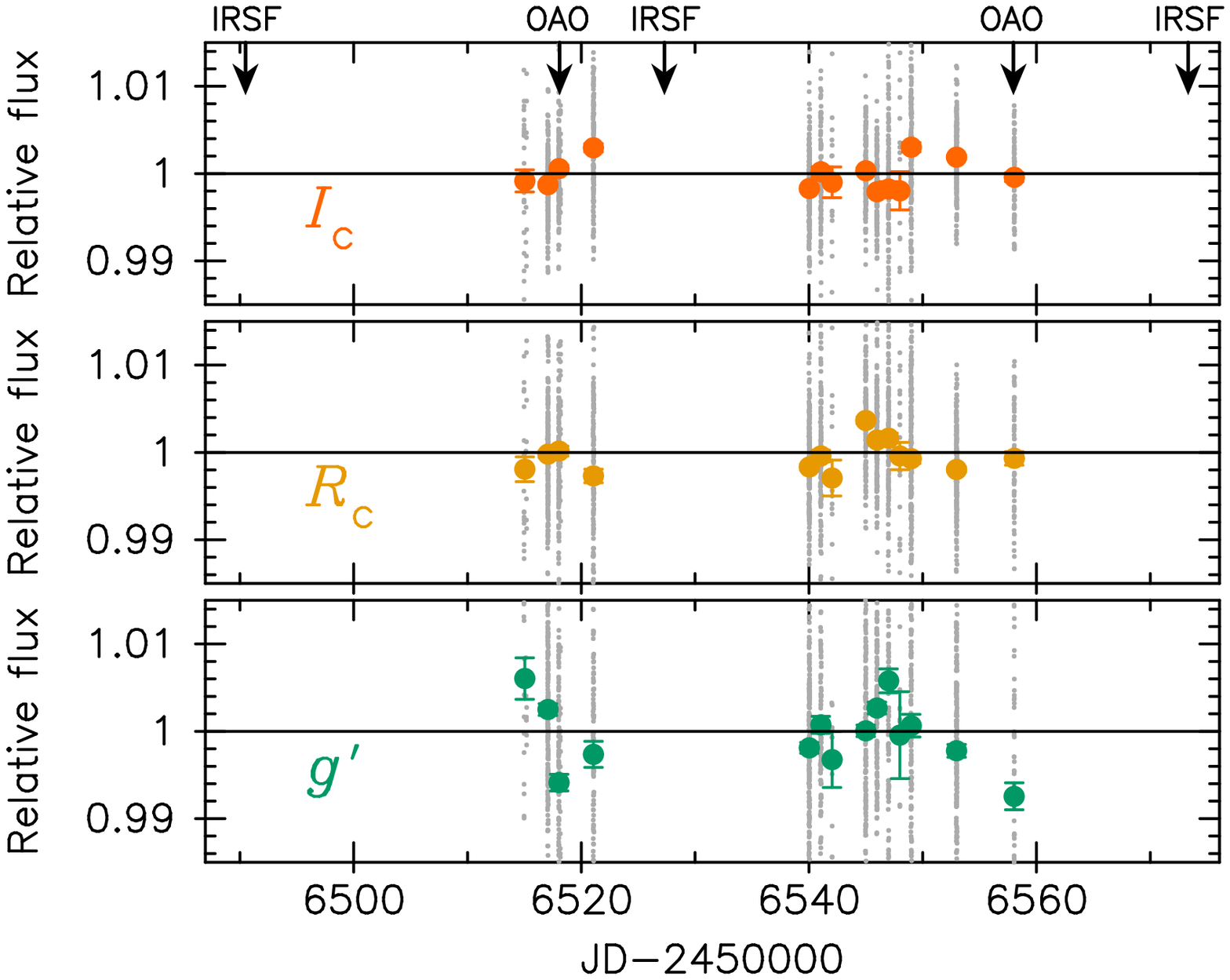}
\caption{Forty-three-day-long light curves of WASP-80 observed with the MITSuME telescope in $I_\mathrm{c}$, $R_\mathrm{c}$, and $g'$ bands (top, middle, and bottom panels, respectively). The gray dots are unbinned data points excluding those that fall in any transits, whereas  the large points indicate nightly binned data. The error bars are calculated as RMS of nightly unbinned data divided by square of the number of data points. The nights when the transit observations were conducted at IRSF and OAO are indicated as arrows in the top panel.\label{fig:lc_longterm}}
\end{center}
\end{figure}

For the 43 day long photometric monitoring data gathered by the MITSuME telescope, we perform aperture photometry 
in the same way as in Sections \ref{sec:reduction} and \ref{sec:prepare_lc}. After eliminating the data points fallen in any transit events, we correct the systematics on the respective light curves by fitting them with Equations (1) and (2) fixing $F_\mathrm{tr}$=1 and using $\{{\bf X}\}$=\{$z$, $\Delta x$, $\Delta y$\}. The resultant light curves with nightly binned data points are shown in Figure \ref{fig:lc_longterm}, in which the error bars are calculated as rms of the unbinned data on each night divided by the square of the number of data points.

\section{Discussion}
\subsection{Atmospheric Properties}
\label{sec:spectrum}

\begin{figure*}
\begin{center}
\includegraphics[width=15cm]{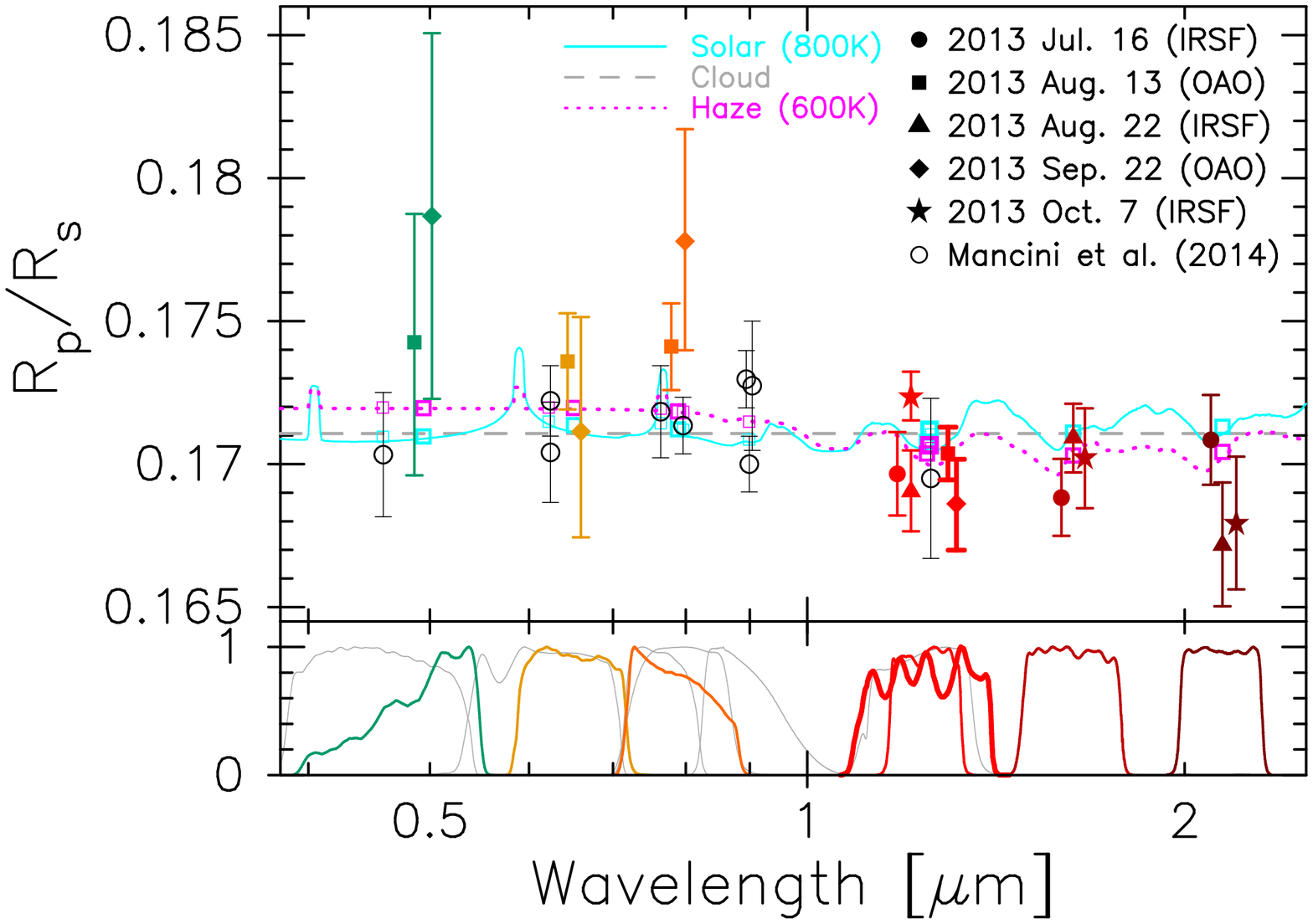}
\caption{(Top) Observed and theoretical transmission spectra of WASP-80b. The $R_\mathrm{p}/R_\mathrm{s}$ values obtained in this work are indicated as filled marks, where different marks are for the data obtained on different nights as indicated at the upper right in the panel. The $J$-band data from ISLE are indicated as bold lines for ease of recognizing. The open circles indicate the data from \citet{2014A&A...562A.126M}, which include re-analysis of the data presented in \citet{2013A&A...551A..80T}. Data points are slightly shifted in the horizontal direction for display when there are multiple observations at one band pass.  The solid, dashed, and dotted lines show theoretical spectra for solar-abundance (800K), cloudy, and hazy (600K) atmospheres, respectively (see text for details). Note the theoretical spectra shown are smoothed for clarity. The open squares indicate the integrals of the theoretical spectra for the respective band passes. (Bottom) Transmission curves for the respective filters, where the bold, normal, and thin lines are for the ISLE/$J$-band filter, the other filters used in this work, and those used in \citet{2014A&A...562A.126M}, respectively.\label{fig:RpRs}}
\end{center}
\end{figure*}

In this section, we discuss the dependence of 
 $R_\mathrm{p}/R_\mathrm{s}$ on wavelength that could arise from the atmospheric properties of WASP-80b.
In Figure \ref{fig:RpRs}, we plot the observed $R_\mathrm{p}/R_\mathrm{s}$ values derived in Section \ref{sec:mcmc} as a function of wavelength, along with those derived by \citet{2014A&A...562A.126M}. 
At each band, the measured values from different transits are consistent with each other within 2$\sigma$ uncertainties. 
Especially in the $J$ band, we have now a total of six observations of $R_\mathrm{p}/R_\mathrm{s}$ including the one from \citet{2014A&A...562A.126M}, and they are mostly consistent with each other, indicating the correctness of our methodology and the smallness of systematics. 
We note that the effect of stellar intrinsic variability on measured $R_\mathrm{p}/R_\mathrm{s}$ is negligible, as will be discussed in Section \ref{sec:stellar_variability}.

In order to search for atmospheric features in the observed $R_\mathrm{p}/R_\mathrm{s}$ spectrum, we first compare the overall observed data with two possible model spectra; one is from an  atmospheric model with the solar abundances and the other is from an atmospheric model with opaque clouds.  For the former model, we simulate the model spectrum assuming a solar abundance atmosphere and a temperature of 800~K, which corresponds to the equilibrium temperature with albedo of 0.1 \citep{2013A&A...551A..80T}, as described in the Appendix; for the latter model, we approximate it simply as a flat line assuming that the atmosphere is thoroughly opaque in the wavelength range from optical to NIR. Note that our motivations for comparing with the cloudy model come from the fact that  flat transmission spectra compatible with the presence of thick cloud layers have been recently observed for several other low-temperature exoplanets such as GJ1214b \citep{2014Natur.505...69K} and GJ436b \citep{2014ApJ...785..126K}, as well as the fact that \citet{2014A&A...562A.126M} reported that their observed spectrum of WASP-80b is consistent with a flat line. 
We then fit the two model spectra to the data; for both cases, the number of free parameters is one: the radius of planetary disk that blocks the incident stellar radiation completely, $R_0$. 

In the fitting process, we create a theoretical $R_\mathrm{p}/R_\mathrm{s}$ spectrum for each $R_0$ using Equation (A2) and the fixed $R_\mathrm{s}$ value of 0.63 $R_\odot$ \citep{2013A&A...551A..80T}, and integrate the spectrum over each filter's pass band to compare it with the observational data.
The best-fitted models for the solar abundance atmosphere and the cloudy atmosphere are shown, respectively, by the solid cyan and dashed gray lines in Figure \ref{fig:RpRs}. As a result, we find that the solar abundance and cloudy models give the minimum-$\chi^2$ values of 35.4 and 34.7, respectively, for the degrees of freedom (dof) of 25. These values indicate that the two models are both largely consistent with the data at  the discrepancy level of 1.7$\sigma$. When we discard the MITSuME data, which were obtained with a relatively small-aperture telescope and might contain relatively large unknown systematics, the solar abundance and cloudy models fit the remaining data with $\chi^2$/dof=24.3/19 and 22.7/19, respectively, reducing the discrepancy levels to 1.3 and 1.1$\sigma$. Therefore, we cannot rule out these two models from the current observational data.

On the other hand, we also find that the observed $R_\mathrm{p}/R_\mathrm{s}$ in the optical region is marginally larger than that in the NIR region; the weighted mean of the observed $R_\mathrm{p}/R_\mathrm{s}$ in the optical  ($\lambda < 1 \mu$m) and NIR ($\lambda > 1 \mu$m) regions  are 0.17193 $\pm$ 0.00041 and 0.17029 $\pm$ 0.00039, respectively, having a 2.9-$\sigma$ discrepancy.
As discussed in Section \ref{sec:intro}, because the equilibrium temperature of WASP-80b is about 800~K or less, photochemically produced hydro-carbon haze like tholin may exist in the atmosphere \citep[e.g.,][]{2013ApJ...775...80F}. 
If so, the observed spectral rise in the optical region could be explained by the existence of tholin haze in the planetary atmosphere.

Motivated by this possibility,
we model the transmission spectrum of a hazy atmosphere with the solar abundances and compare it with the observed data. 
The haze layer is characterized by four parameters that include the particle size $a_\mathrm{haze}$, the number density $n_\mathrm{haze}$, and the pressures at the top and bottom of the haze layer, which are denoted by $P_\mathrm{top}$ and $P_\mathrm{bot}$, respectively. 
Values of $P_\mathrm{top}$ and $P_\mathrm{bot}$ are chosen as described in the Appendix; in the case of GJ1214b, the method of choice is confirmed to yield values of $P_\mathrm{top}$ and $P_\mathrm{bot}$ that are consistent with the result from \citet{2013ApJ...775...33M}.
We assume $a_\mathrm{haze} = 0.04$~$\mu$m, which is the typical size of haze particles in Titan's atmosphere \citep{2009Icar..204..271T}. 
We then search for the best-fit hazy model by changing $n_\mathrm{haze}$ every one order of magnitude from 10 to $1\times 10^6$~cm$^{-3}$, as well as letting $R_0$ be free.  
As a result, we find that $n_\mathrm{haze} = 1\times10^4$~cm$^{-3}$ gives a minimum-$\chi^2$ value of 29.3 with dof = 24 (in this case, the number of free parameters is two) for temperature of 800~K, which means that the discrepancy level is 1.3$\sigma$. A comparable fit to the data without the MITSuME data gives $\chi^2$/dof = 22.7/18, or 0.99$\sigma$. These values are slightly better than those for the above two models of the haze-free solar-abundance atmosphere and the cloudy atmosphere.

In reality, however, we may have to consider temperature lower than 800~K. 
The 800~K corresponds to the globally averaged equilibrium temperature of WASP-80b with an albedo of 0.1. 
Since WASP-80b is likely to be tidally locked, the limb of the planetary disk that we observe may be much cooler than 800~K, provided the atmospheric heat redistribution is inefficient. 
Also, the high-altitude haze may block incident stellar flux from reaching the deep atmosphere. 
Thus, we consider 600~K, as an example of a moderately warm atmosphere to simulate the transmission spectrum which is shown by the dotted magenta line in Figure~\ref{fig:RpRs}. 
In this case, the best-fit model (with $n_\mathrm{haze}= 1\times10^4$~cm$^{-3}$) results in a $\chi^2$/dof of 26.8/24, giving a discrepancy level of only 1.0$\sigma$. These statistical values decrease to $\chi^2$/dof=19.5/18 and 0.92$\sigma$ for the case without the MITSuME data. These results indicate
 that the hazy atmosphere model with temperature of 600~K is rather consistent with the observed data, relative to the above three models. 
In contrast, we also find that the haze-free solar abundance atmosphere of 600~K yields a large $\chi^2$/dof  of 44.2/25 and 29.7/19 for the data with and without the MITSuME data, respectively.
In Table \ref{tbl:model_fitting}, we summarize the statistical results of the model fittings discussed above.

A more extensive search for the best-fit model is beyond the scope of this study because the observed data have uncertainties that are too large and wavelength resolutions that are too low. 
Also, we need more detailed treatment concerning the role of haze on the atmospheric temperature and the heat redistribution in the atmosphere, which will be left to future studies. 
Nevertheless, we confirm that at least one atmospheric model with haze can explain the observed data well. 
Thus, the relatively large $R_\mathrm{p}/R_\mathrm{s}$ in the optical region detected by our observation has raised another possibility of the presence of haze in the atmosphere. 
Higher-precision and higher-wavelength-resolution observations are desired to further shed light on this possibility.

\begin{deluxetable}{cccccc}
\tablewidth{8.5cm}
\tablecaption{Statistical Results of Model Fitting to the Transmission Spectrum \label{tbl:model_fitting}}
\tablehead{
\colhead{} & \colhead{Solar} & \colhead{Cloud} & \colhead{Haze } &  \colhead{Haze} & \colhead{Solar}\\[1pt]
& (800 K) & & (800 K) & (600 K) & (600 K)
}
\startdata
\multicolumn{6}{c}{\it All Data}\\[1pt]
$\chi^2$/dof & 35.4/25 & 34.7/25 & 29.3/24 &  26.8/24 & 44.2/25\\[1pt]
Discrepancy [$\sigma$] & 1.7 & 1.7 & 1.3 & 1.0 & 2.7 \\[5pt]
\multicolumn{6}{c}{\it w/o MITSuME Data}\\[1pt]
$\chi^2$/dof & 24.3/19 & 22.7/19 & 20.2/18 &  19.5/18 & 29.7/19\\[1pt]
Discrepancy [$\sigma$] & 1.3 & 1.1 & 0.99 & 0.92 & 1.9

\enddata
\end{deluxetable}

\subsection{Transit Timings}

Due to the multi-epoch observations, we are able to refine the transit ephemeris as well as search for transit timing variations (TTVs) that would be caused by additional perturbing planets \citep[e.g.,][]{2010Sci...330...51H}. The latter is particularly of interest because  warm Jupiters may have higher probability of having TTV-causing neighboring planets compared to hot Jupiters, most of which are known to  be solitary \citep[e.g.,][]{2012PNAS..109.7982S}.

\begin{figure}
\begin{center}
\includegraphics[width=8.5cm]{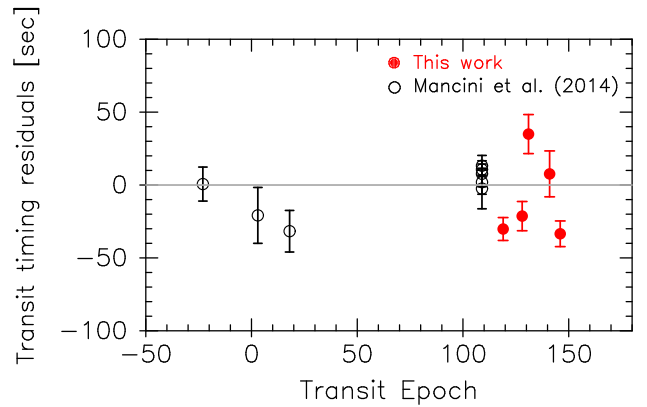}
\caption{Residuals of the observed mid-transit times from the liner ephemeris calculated by fitting all the data shown here. The latest five data points are from this work, and others are from \citet{2014A&A...562A.126M} including the re-analysis data of the light curves presented in \citet{2013A&A...551A..80T}. \label{fig:TT}}
\end{center}
\end{figure}

Using the mid-transit times of the five transits measured in Section \ref{sec:mcmc} as well as those from the previous works, we refine the transit ephemeris of WASP-80b as $T_\mathrm{c}\ \mathrm{(BJD_{TDB})} =$ 2456125.417574 (86) + 3.06785952\ (77) $\times E$, where $E$ is the relative transit epoch and the numbers in parentheses indicate the uncertainties written to the last two significant digits.
The residuals of the observed transit timings from the above ephemeris are shown in Figure \ref{fig:TT}.
The $\chi^2$ value for the linear fit is 59.0 for the degrees of freedom of 11, indicating that a liner function does not fit the data well. This in principle could  be due to perturbations from an additional object in the planetary system, however, it could also be due to small-number statistics and/or unknown systematics as well \citep[e.g.][]{2013AJ....146..147M,2013MNRAS.430.3032B,2012MNRAS.420.2580S,2012ApJ...748...22H}. In addition, we cannot see any plausible periodicity nor large amplitude exceeding $\sim$50 s that are usually seen in the detection cases \citep[e.g.][]{2013ApJS..208...16M}. Therefore, we do not claim a detection of TTVs due to a third body at this time.

\subsection{Stellar Variability}
\label{sec:stellar_variability}

To check for the existence/absence of stellar intrinsic variability that causes systematic offsets on the observed  $R_\mathrm{p}/R_\mathrm{s}$, we investigate the 43 day long light curves created in Section \ref{sec:longlc}.
The $\chi^2$ values of the 14 nightly binned data with respect to a constant fit are 212.7, 181.8, and 137.4 for $g'$, $R_\mathrm{c}$, and $I_\mathrm{c}$ bands, respectively, suggesting that the host star's brightness could significantly vary over time. However, the data points in different bands are not correlated each other, with the correlation coefficients between $I_\mathrm{c}$ and $R_\mathrm{c}$, $R_\mathrm{c}$ and $g'$, and $g'$ and $I_\mathrm{c}$ are $-0.26$, 0.29, $-0.32$, respectively. This fact indicates that the observed dispersions are due to systematics rater than astrophysical origin.
In addition, no periodic variation or linear trend can be seen in any of the three light curves. Therefore, we do not detect any star-spot-induced periodic variability with the semi-amplitude larger than 0.25\%, 0.3\%, and 0.7\% for $I_\mathrm{c}$, $R_\mathrm{c}$, and $g'$ bands, respectively, during the observed period range of 43 days. Even if the observed dispersions were of astrophysical origins, the maximum variability of 0.7\% in the $g'$ band would change the observed $R_\mathrm{p}/R_\mathrm{s}$ values  by only 0.07\%, or 0.0001 in units of $R_\mathrm{p}/R_\mathrm{s}$, which is negligible compared to the uncertainties of $R_\mathrm{p}/R_\mathrm{s}$.

The non-detection of significant stellar variability is in line with the report of \citet{2013A&A...551A..80T}, who did not detect $>$1 mmag rotational variability from wide-band ($V$+$R$) photometric data 
of the WASP transit survey as well as from multi-epoch observations with a 60 cm telescope. We therefore confirm from the multi-band observations that the host star WASP-80 does not show spot-induced large ($>$ a few mmag) periodic variations. 
In addition, so far no spot-crossing event has been observed during any of the transits observed in this work or in previous works (nine transits in total), implying that there are not many or large spots distributed on the stellar surface.
On the other hand, \citet{2013A&A...551A..80T} pointed out that the star could be young because of its high projected stellar rotational velocity ($v \mathrm{sin} i_\mathrm{s} = 3.55 \pm 0.33$ km s$^{-1}$), depletion of lithium, and the presence of Ca $H$+$K$ emission, although there is also counter-evidence that the galactic dynamical velocities are low. In addition, \citet{2014A&A...562A.126M} detected strong Ca $H$+$K$ emission lines, indicating that the star has strong magnetic activity. If the star is truly young and active, the star would be expected to be heavily spotted. If so, a possible scenario would be that either many small spots are widely distributed on the stellar surface \citep{2014A&A...562A.126M}, or that the stellar inclination $i_\mathrm{s}$ is very small such that the polar region of the stellar surface is always facing us. The latter idea is particularly consistent with the fact that the projected spin-orbit angle $\beta$ was measured as a significantly non-zero value ($\pm75^{\circ} \pm 4$), assuming $V \mathrm{sin} i_\mathrm{s}$ = $v \mathrm{sin}  i_\mathrm{s}$, where $V \mathrm{sin} i_\mathrm{s}$ and $v \mathrm{sin}  i_\mathrm{s}$ are the projected stellar-rotation velocities measured from the Rossiter-McLaughlin effect and spectral equivalent width, respectively \citep{2013A&A...551A..80T}. In this case, the star should be rotating very fast, possibly less than a few days. Photometric monitoring of the star with much higher precision may help to measure the rotational period and test this possibility.

\section{Summary}
In this paper, we report multi-color, multi-epoch observations of the transiting warm Jupiter WASP-80b, which is suitable for studying an exoplanetary atmosphere with low temperature. We observed five primary transits of this planet using three instruments and seven different filters. Consequently, we obtained 17 independent transit light curves in six colors covering from optical to near infrared (NIR) wavelength regions. We compare the observed $R_\mathrm{p}/R_\mathrm{s}$ values including those from previous works with two model spectra, one is from a solar-abundance atmospheric model and the other is from a thick cloud one. As a result, we find that the observed data are largely consistent with both the two models at 1.7$\sigma$. Therefore, we cannot rule out these two models from the current observations. 

On the other hand, we also find that the observed $R_\mathrm{p}/R_\mathrm{s}$ in optical is marginally larger than that in NIR at 2.9$\sigma$ significance, possibly indicating the existence of haze in the planetary atmosphere. 
We compare the data with theoretical spectra for a solar abundance but hazy atmosphere, and find that a model with the equilibrium temperature of 600K fits the data at 1.0$\sigma$, indicating that the hazy atmospheric model can explain the observed data well. To confirm or reject this possibility, further higher-precision and higher-spectral-resolution observations are required.

We also search for transit timing variations from totally 13 timing data (9 transits) from this work and previous work.
As a result, we do not find any periodic timing variation nor timing excess larger than 50 s from a linear ephemeris, meaning that there is no considerable sign of additional neighboring planet at this time. 

In addition, we conducted 43 day long photometric monitoring of the host star in $g'$, $R_\mathrm{c}$, and $I_\mathrm{c}$ bands, resulting in a non-detection of significant brightness variations. Combining with the fact that no spot-crossing event is observed in the five transits, we confirm the findings of \citet{2013A&A...551A..80T} and \citet{2014A&A...562A.126M} that the host star appears quiet for spot activities despite indications of strong chromospheric activities. This odd consequence could be explained if the host star has a polar-on orbit. This possibility can be tested by measuring the stellar rotational period by monitoring the stellar brightness with much higher precision.

\acknowledgments
We thank L. Mancini for kindly providing filter information. A.~F. thanks M.~Kuzuhara for meaningful discussion about atmospheric study of gas giants.
Y.~K. and M.~I. thank N.~Iwagami and Y.~Ito for fruitful discussions about modeling of the transmission spectra. 
This work is partially supported by the Optical \& Near-Infrared
Astronomy Inter-University Cooperation Program from the Ministry of Education, Culture, Sports, Science and Technology  of Japan (MEXT).
N.~N. acknowledges support by the NAOJ Fellowship,
Inoue Science Research Award, and
Grant-in-Aid for Scientific Research (A) (No.~25247026) from the MEXT. S.~N. and M.~T. are supported by the Japan Society for the Promotion of Science
 (JSPS) Grant-in-Aid: Young Scientists (A) (No.~25707012) and No.~22000005, respectively.
T.H., Y.H., and K.K. are supported by Grants-in-Aid for JSPS Research Fellows: No. 25-3183, 25000465, and 26-11515, respectively.

\appendix
\section{Outline of Transmission Spectrum Modeling}

The theoretical value of the transit radius at wavelength $\lambda$, $R_\mathrm{p, th} (\lambda) $, is calculated as
\begin{equation}
 \left[ \frac{R_\mathrm{p, th} (\lambda)}{R_\mathrm{s}} \right]^2 = 
 \frac{1}{\pi R_\mathrm{s}^2} 
 \int_0^{R_\mathrm{s}} 
 \left\{1 - e^{-\tau (r, \lambda)} \right\}
 \cdot 2 \pi r dr,
\end{equation}
where $R_\mathrm{s}$ is the host star's radius, $r$ is the planetocentric distance, and $\tau$ is the chord optical thickness \citep[see e.g.][]{2014A&A...562A..80K}.  
We assume that the planetary disk of radius $R_0$ blocks the incident stellar radiation completely, which means 
\begin{equation}
 \left\{ R_\mathrm{p, th} (\lambda) \right\}^2 = R_0^2 + 
 \int_{R_0^2}^{R_\mathrm{s}^2} 
 \left\{1 - e^{-\tau (r, \lambda)} \right\} dr^2.
\end{equation}
In this study, we define $R_0$ as the planetocentric distance at which the atmospheric pressure is 10~bar.  
Since the pressure profile in the atmosphere is unknown in advance, we treat $R_0$ as a free parameter while searching for the best-fit model by $\chi^2$ analysis.
The atmosphere is assumed to be in hydrostatic equilibrium and isothermal for simplicity; $R_\mathrm{p, th}$ is known to be less sensitive to atmospheric pressure-temperature profiles \citep{2010ApJ...716L..74M,2012ApJ...756..176H}.  
We also assume that the element abundances are radially constant and calculate chemical-equilibrium molar fractions of molecules at each altitude with the Gibbs free energy data from NIST-JANAF Thermochemical Tables \citep{JANAF}. We determine the element abundance ratios of the solar abundance atmosphere from \citet{2003ApJ...591.1220L}.

As for the sources of radiative extinction, we consider line absorption by H$_2$, H$_2$O, CH$_4$, CO, CO$_2$, NH$_3$, N$_2$, Na, and K gases, and collision-induced absorption by H$_2$-H$_2$, and H$_2$-He for the solar abundance model, and additionally scattering by haze particles for the hazy atmospheric models. 
We take line data for those gaseous molecules except Na and K from HITRAN2012 \citep{2013JQSRT.130....4R} and those for Na and K from \citet{1992RMxAA..23...45K}, and calculate the absorption cross sections for those gases with the Voigt profile \citep[e.g.,][]{1989AtmosphericRadiation}. In practice, we use the geometric mean of the wavelength-dependent cross sections over a range with a wavenumber width of 3.3~cm$^{-1}$. 
The cross sections due to the collision-induced absorption are taken from HITRAN2012. 

For haze particles, we assume hydrocarbon haze, which is often called tholin. 
Taking its complex indices of refraction from \citet{1984Icar...60..127K}, we calculate its extinction coefficients based on the Mie theory, using HITRAN-RI program in HITRAN2012. 
The haze layer is characterized by four parameters that include the particle size $a_\mathrm{haze}$, the number density $n_\mathrm{haze}$, and the pressures at the top and bottom of the haze layer which are denoted by $P_\mathrm{top}$ and $P_\mathrm{bot}$, respectively. 
Hydrocarbon haze is usually produced from CH$_4$. CH$_4$ is the major C-bearing molecule in the lower atmosphere, while CO is dominant in the upper atmosphere. 
Thus, hydrocarbon haze should appear around the altitude where CH$_4$ changes to CO. 
According to recent simulations of photochemistry in the atmosphere of GJ1214b by \citet{2013ApJ...775...33M}, the precursor molecules such as C$_2$H$_2$ forms at such an altitude and is distributed in the region that ranges over 1-2 orders of magnitude in pressure. 
In this study, we calculate the equilibrium composition to find the altitude at which CH$_4$ is equal in mole fraction to CO. 
While our calculation does not include photo-chemical effects and assumes the isothermal structure, we have checked that our calculated altitude for GJ1214b is similar with that from \citet{2013ApJ...775...33M}. 
The calculated pressures at that altitude for WASP-80b are approximately $1 \times 10^{-4}$~bar and $1 \times 10^{-2}$~bar for temperatures, $T$, of 600~K and 800~K, respectively. 
Thus, we assume that 
$P_\mathrm{top} = 1 \times 10^{-5}$~bar and $P_\mathrm{bot} = 1 \times 10^{-3}$~bar for $T = 600$~K and that 
$P_\mathrm{top} = 1 \times 10^{-3}$~bar and $P_\mathrm{bot} = 0.1$~bar for $T = 800$~K.

As for the particle size, we assume $a_\mathrm{haze}$ = 0.04~$\mu$m, which is the typical size of haze particles observed in Titan's atmosphere \citep{2009Icar..204..271T}. 
Larger particles with $a_\mathrm{haze} \gtrsim 0.1$~$\mu$m would be incompatible with the spectral feature such that the transit radius is larger in the optical region than in the NIR region. 
For $a_\mathrm{haze} \lesssim 0.1$~$\mu$m, the Rayleigh scattering by haze particles determines the spectrum in the optical region. 
For an appropriate choice of $n_\mathrm{haze}$, the Rayleigh slope would be consistent with the spectral feature that we have observed. 
Because different sets of $a_\mathrm{haze}$ and $n_\mathrm{haze}$ yield similar spectral features, we regard $n_\mathrm{haze}$ as a free parameter in this study.
The number density $n_\mathrm{haze}$ is assumed to be between 10~cm$^{-3}$ and $1 \times 10^6$~cm$^{-3}$, which is an expected range in Titan's atmosphere \citep{2007ApJ...661L.199L} and hydrogen-rich atmospheres of warm exoplanets \citep[e.g.][]{2013ApJ...775...33M}, although its exact value is uncertain.

\vspace{10pt}


\end{document}